\pdfoutput=1

\documentclass[11pt,twoside,a4paper,cmspaper,final,collab]{cms-tdr}
\def\svnVersion{175878}\def\cmsCernNo{2012-164}\def\cmsCernDate{\today}\def\cmsMessage{Submitted to the Journal of High Energy Physics}

\begin{document}\cmsNoteHeader{EXO-11-067}

\hyphenation{had-ron-i-za-tion}
\hyphenation{cal-or-i-me-ter}
\hyphenation{de-vices}

\RCS$Revision: 149663 $
\RCS$HeadURL: svn+ssh://svn.cern.ch/reps/tdr2/papers/EXO-11-067/trunk/EXO-11-067.tex $
\RCS$Id: EXO-11-067.tex 149663 2012-09-28 06:20:10Z hlliu $
\providecommand{\PSGczDo}{\ensuremath{\widetilde{\chi}^0_1}\xspace} 
\cmsNoteHeader{EXO-11-067} 
\title{Search for new physics with long-lived particles decaying to photons and missing energy in pp collisions at $\sqrt{s}=7\TeV$}

\date{\today}

\abstract{
A search is performed for long-lived neutral particles decaying into a photon and invisible particles. An example of such a signature is the decay of the lightest neutralino with nonzero lifetime into a gravitino and a photon in gauge-mediated supersymmetry, with the neutralino as the next-to-lightest supersymmetric particle and the gravitino as the lightest. The search uses events containing photons, missing transverse energy, and jets. The impact parameter of the photon relative to the beam--beam collision point can be reconstructed using converted photons. The method is sensitive to lifetimes of the order of 0.1 to 1\unit{ns}. The data sample corresponds to an integrated luminosity of 2.23\fbinv in pp collisions at $\sqrt{s} = 7\TeV$, recorded in the first part of 2011 by the CMS experiment at the LHC. Cross-section limits are presented on pair production for such particles, each of which decays into a photon and invisible particles. The observed 95\% confidence level limits vary between 0.11 and 0.21\unit{pb}, depending on the neutral particle lifetime.
}

\hypersetup{%
pdfauthor={CMS Collaboration},%
pdftitle={Search for new physics with long-lived particles decaying to photons and missing energy in pp collisions at sqrt(s) = 7 TeV},%
pdfsubject={CMS},%
pdfkeywords={CMS, physics, software, computing}}

\maketitle 

\section{Introduction}
New heavy particles with long lifetimes are predicted in many models of physics beyond the standard model (SM), including models with gauge-mediated supersymmetry breaking (GMSB)~\cite{Ambrosanio:1996jn, Dimopoulos:1996yq, Giudice:1998bp, Carpenter:2008he, Rajaraman:2009ga, Abel:2009ve, Meade:2009qv,  Meade:2010ji, Ruderman:2011vv} and certain classes of hidden valley models~\cite{Strassler:2006im}. Such particles may be neutral and decay into photons and invisible particles, which escape detection. Their lifetime is essentially a free parameter of the models, which motivates searches sensitive to a wide range of potential lifetimes. For sufficiently long lifetimes, of order of 1 to 10\unit{ns}, measurement of the shower direction by the D0 experiment~\cite{Abazov:2008zm} or time-of-flight by the CDF experiment~\cite{Abulencia:2007ut} with an electromagnetic calorimeter can be used to identify such decays, and limits have been obtained by using such techniques at the Tevatron.

In this paper, we present a new analysis method that is sensitive to neutral particles with lifetimes of the order of 0.1 to 1\unit{ns}. In a fraction of events,  photons undergo conversion into \Pep\Pem~ pairs in the material of the LHC beam pipe or the CMS charged-particle tracking system, providing a clear experimental signature. The tracks of the electrons can be precisely reconstructed and used to calculate the photon trajectory. In particular, the photon impact parameter with respect to the primary vertex can be determined.

\begin{figure}[h!]
  \begin{center}
    \includegraphics[width=0.62\textwidth]{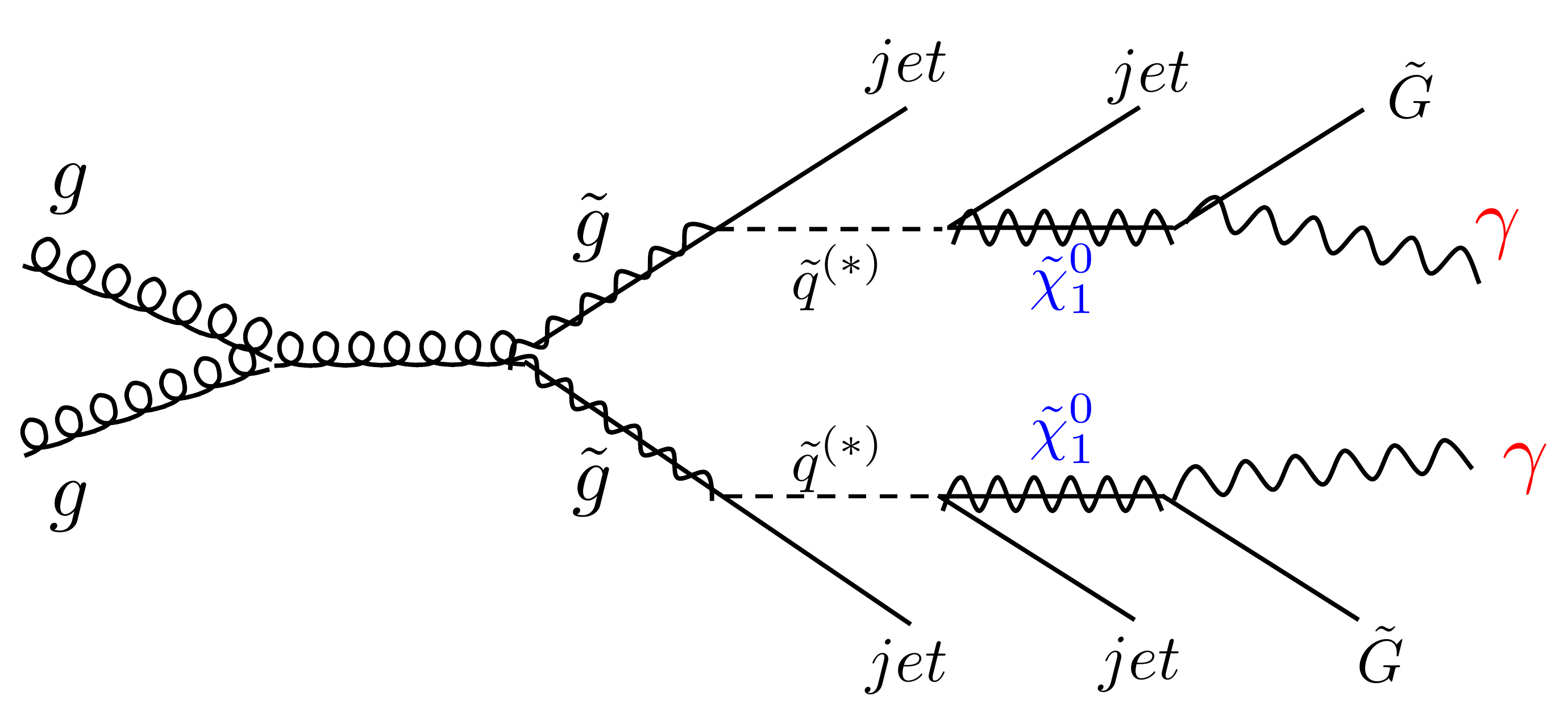}
    \caption{Example diagram of $\PSGczDo$ pair production and $\PSGczDo\rightarrow \gamma + \PXXSG$ decay.}
    \label{fig:feyman}
  \end{center}
\end{figure}

We search for diphoton events with at least one converted photon having a significant impact parameter produced in association with missing transverse energy. Models with gauge-mediated supersymmetry breaking, e.g., general gauge-mediation (GGM), are used to demonstrate the kinematics of the production and decay. Figure~\ref{fig:feyman} shows one of the production and decay diagrams. Assuming that R-parity is conserved~\cite{Farrar:1978xj}, SUSY particles are produced in pairs and decay into SM particles and the lightest neutralino ($\PSGczDo$). The neutralino decays into a photon and a gravitino ($\PXXSG$), the lightest SUSY particle in this model, which escapes the detector, leading to apparent missing transverse energy \MET. Moreover, in many models, the $\PSGczDo$ is produced in association with high transverse momentum (\pt) jets. We consider neutralino proper decay lengths between $c\tau =$ 2 and 25\unit{cm}, corresponding to a lifetime $\tau$ on the order of 0.1 to 1\unit{ns}.

The data sample was collected in 2011 by the Compact Muon Solenoid (CMS) experiment at the LHC from proton--proton collisions at a center-of-mass energy of 7 TeV, corresponding to an integrated luminosity of $2.23\pm 0.05\fbinv$. A detailed description of the CMS detector can be found elsewhere~\cite{JINST}. The detector's central feature is a superconducting solenoid providing a 3.8\unit{T} axial magnetic field along the beam direction. Charged particle trajectories are measured by a silicon pixel and strip tracker system, covering $0 \leq \phi \leq 2\pi$ in azimuth and $\abs{\eta} < 2.5$, where pseudorapidity is defined as $\eta=-\ln\tan\theta/2$, and $\theta$ is the polar angle with respect to the counterclockwise beam direction. The amount of material crossed by a particle traversing the tracker volume from the interaction region varies from 0.4 to more than 2 radiation lengths, depending on the particle direction. A lead-tungstate crystal electromagnetic calorimeter (ECAL) and a brass--scintillator hadron calorimeter (HCAL) surround the tracker volume. For the barrel and endcap calorimeters ($\abs{\eta} < 3.0$), the modules are arranged in projective towers. Muons are measured with gaseous detectors embedded in the steel return yoke of the magnet. The detector is nearly hermetic with calorimeter coverage up to $\abs{\eta}=5.0$, allowing reliable measurement of \MET.

The analysis strategy consists of selecting events with a diphoton final state with at least one converted photon, and then examining the impact parameter of each single photon for a displaced photon signal. Missing transverse energy and at least two jets are also required. The background is estimated using a low \MET control sample. Upper limits on the cross section for pair production of neutral particles, each of which decays into a photon and invisible particles, are then computed as a function of the neutral particle's lifetime.

\section{Event sample and selections}

The data were recorded using the CMS two-level trigger system. This analysis selects events with at least two photons. A diphoton trigger is required with ECAL transverse energy thresholds \et set to values increasing from 32 (22) to 40\GeV (28\GeV) for the leading (sub-leading) photon, as the instantaneous luminosity rose over the course of the data-taking period. To ensure being on the plateau of the trigger efficiency, the offline analysis selects events with at least two photons with $\et > 45$\GeV (30\GeV) for the leading (sub-leading) photon in the event. The data sample is used both for the selection of signal candidates and for control samples used for background estimation.

The \PYTHIA 6.4 event generator~\cite{Sjostrand:2006za} is used to simulate SUSY signal events. In particular, we generate SUSY GMSB signal datasets
in the SPS8 benchmark model~\cite{Allanach:2002nj} with $\Lambda$ around 100\TeV. Every event in this sample has two neutralinos, each having a mass of 140\GeV. The neutralinos in each dataset have a specific mean lifetime. The signal selection efficiency can depend on the $\PSGczDo$ mass, but for masses around 100\GeV this dependence has a negligible effect in this analysis. Each of the neutralinos decays to a photon and a light gravitino. The CMS detector response is fully simulated using \GEANTfour~\cite{Agostinelli:2002hh}.

Because of the high luminosity during the run, many events had multiple pp interactions leading to a distribution of primary vertices, called ``pile-up''. The pile-up conditions at high luminosity affect the event selections, in particular conversion reconstruction efficiencies. The generated pile-up distribution in the Monte Carlo samples was reweighted to reproduce the 2011 data-taking conditions.

The unconverted and converted photon candidates are reconstructed from clusters of energy in the ECAL. At least one photon candidate with transverse energy $\et>45$\GeV and reconstructed in the ECAL barrel region is required. The transverse distribution of energy in the associated cluster of ECAL crystals must be consistent with that expected from a photon, and the energy detected in the HCAL behind the photon shower cannot exceed 5\% of the ECAL energy. To suppress hadronic jets misreconstructed as photon candidates, we require the latter to be isolated from other particles in the tracker, ECAL, and HCAL. A cone of $\Delta R = \sqrt{(\Delta\eta)^2+(\Delta\phi)^2} = 0.4$ is constructed around the candidate's direction, and the scalar sums of transverse energies of tracks and calorimeter deposits within this cone are determined, after excluding the contribution from the candidate itself. The isolation sums are required to be $\sum E_\text{ECAL} < 0.006 \cdot \et + 4.2$\GeV, $\sum E_\text{HCAL} < 0.0025 \cdot \et + 2.2$\GeV, and $\sum \pt^{\text{tracks}} < 0.001 \cdot \et + 2.0$\GeV, for the ECAL, HCAL, and tracker, respectively, where \et is the transverse energy of the photon candidate. To reject electrons, we require that photon candidates do not have associated hit patterns in the pixel detector consistent with an electron track and not be part of a reconstructed conversion track pair.

Photon-like signals that satisfy the same requirements as photons except for isolation are defined as ``fake photons''. These objects, predominantly jets containing energetic neutral particles such as $\pi^0$s, are used for the background estimation.

Jets are reconstructed from energy deposits in the calorimeters using the anti-\kt clustering algorithm~\cite{Cacciari:2008gp}. The energies of these jets are corrected using the \pt of the charged tracks reconstructed in the tracker~\cite{JME-10-003}. At least two jets are required with $\lvert\eta\rvert\le 2.6$ and leading jet transverse momentum $p_{\rm T1}>80 $\GeV and sub-leading jet transverse momentum $p_{\mathrm{T}2}>50$\GeV.

Missing transverse energy is calculated from calorimeter energy deposits. It is corrected using tracking information for energy missed owing to incomplete calorimeter measurements of muons and charged hadron energies, especially for soft tracks that do not reach the calorimeters~\cite{JME-10-009}. The \MET threshold of 30\GeV discriminates between signal and background.

\section{Photon conversion and the photon impact parameter}
\begin{figure}[h!]
  \begin{center}
    \includegraphics[width=0.36\textwidth]{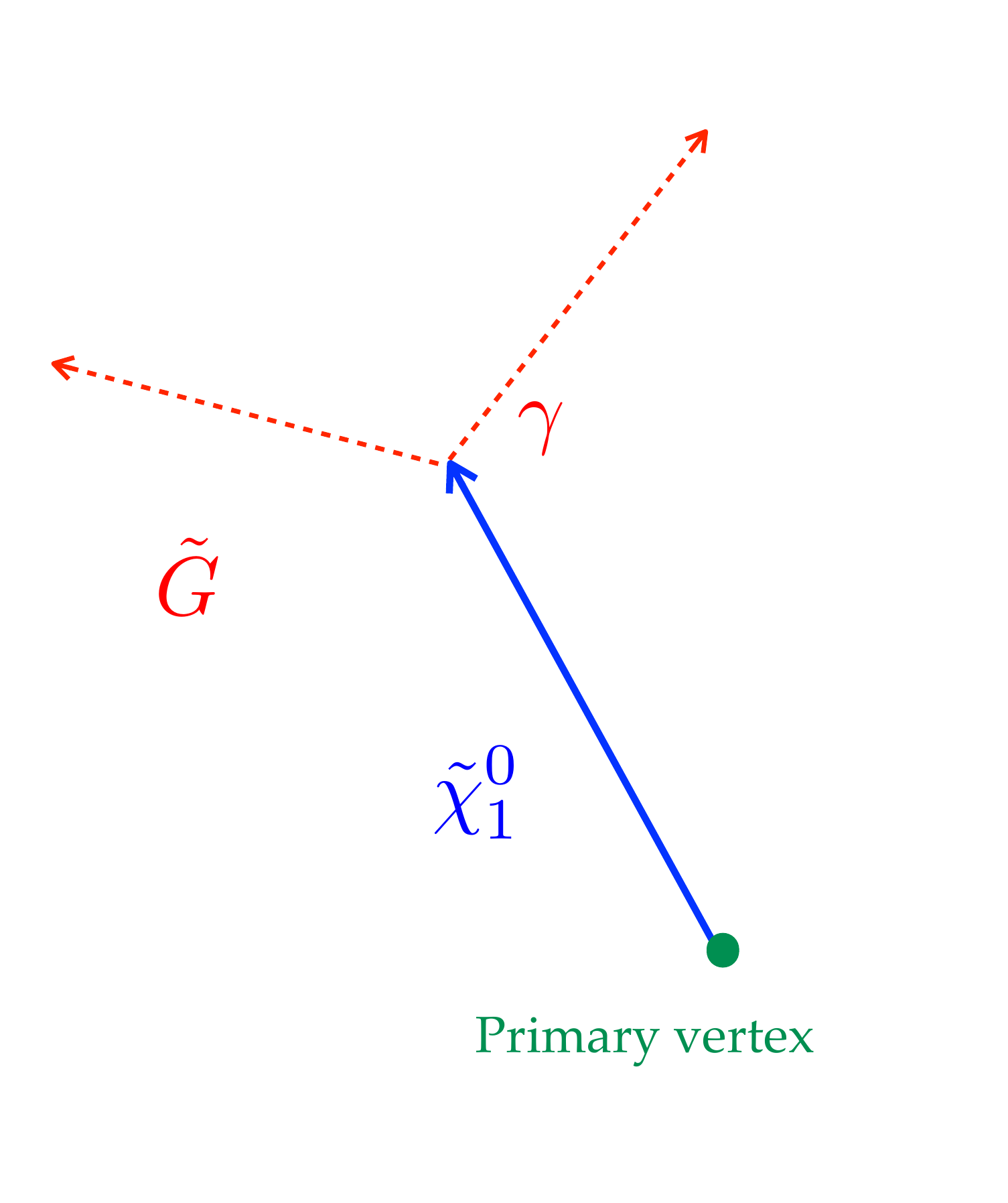}
    \includegraphics[width=0.39\textwidth]{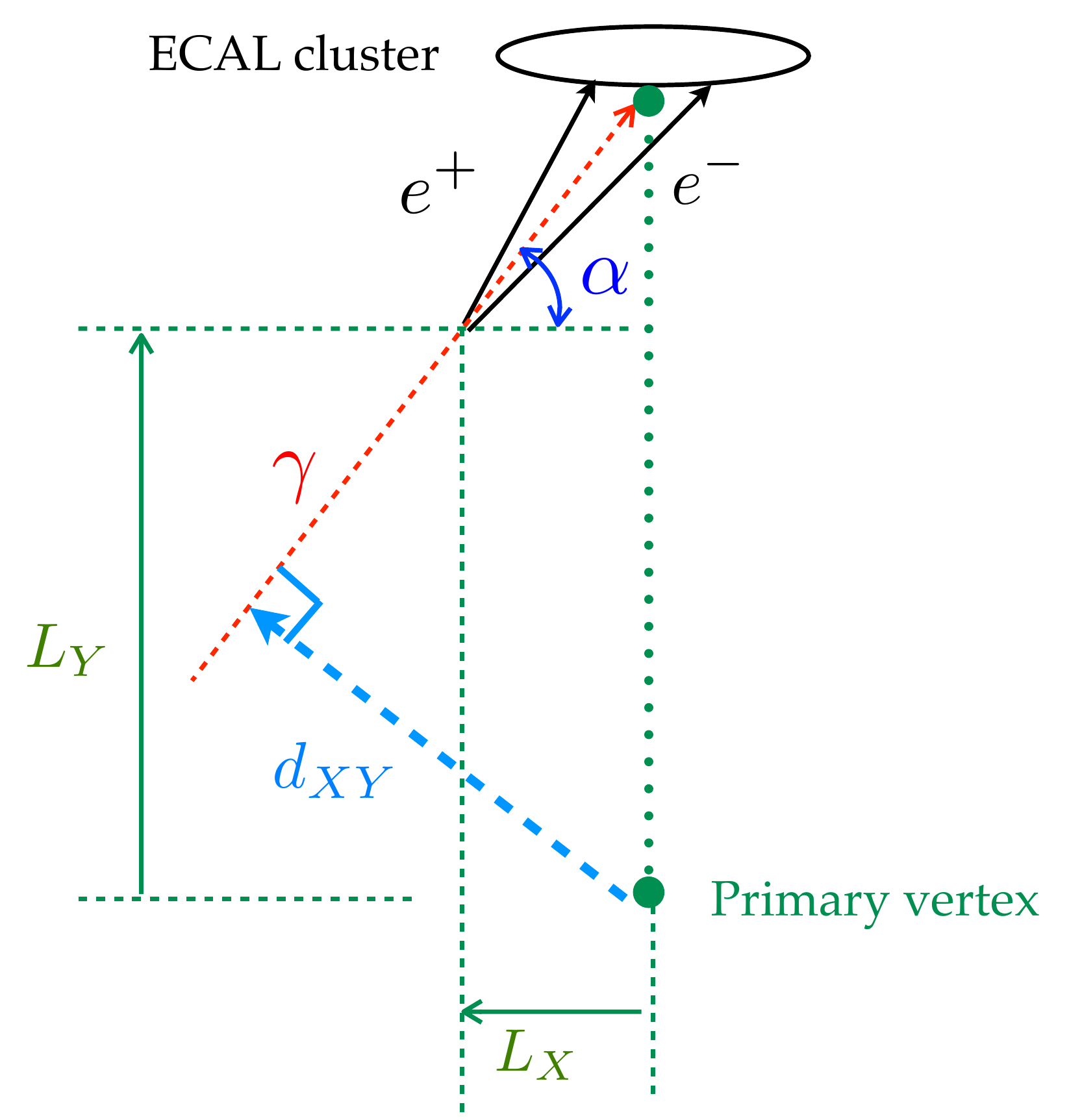}
    \caption{Left: Illustration of $\PSGczDo\rightarrow\gamma +\PXXSG$ decay. Right: In $x$-$y$ plane, a neutralino travels with a finite lifetime and decays into $\gamma +\PXXSG$. The diagram shows the photon converting into an \Pep\Pem~ pair, and the subsequent reconstruction of the impact parameter.}
    \label{fig:decay}
  \end{center}
\end{figure}

The CMS silicon tracker provides robust and precise reconstruction of charged-particle momenta in the high-occupancy environment of LHC collisions. Photon conversions  in the tracker material can be used to obtain the photon direction. By extrapolation along the momentum direction from the conversion vertex back to the beam line, we can calculate the impact parameter of the displaced photons, as shown in Figure~\ref{fig:decay}. If the $\PSGczDo$ has a nonzero lifetime, the decay photon can originate from a displaced secondary vertex rather than close to the primary vertex. The momentum direction determined from conversion tracks for this photon will point away from the primary vertex, giving a nonzero impact parameter (IP), which could be a signature for a long-lived $\PSGczDo$ signal. The conversion reconstruction efficiency is approximately 5\%.

The transverse impact parameter $d_{XY}$ is the distance of closest approach of the photon trajectory to the beam line in the transverse plane to the beam. The longitudinal impact parameter $d_{Z}$ is the distance of closest approach of the photon trajectory to the primary vertex projected on to the $z$ direction. The photon trajectory is defined as a straight line from the conversion vertex along the conversion momentum. The impact parameters $d_{XY}$ and $d_Z$ are expressed in terms of the vector $\vec{\textit{L}}=(L_X, L_Y, L_Z)$ between the conversion point and the primary vertex, and the angle $\alpha$ of the conversion momentum vector $\vec{p}$ projected on to the $x$-$y$ plane as follows:

\begin{equation}
\begin{array}{l}
 \displaystyle d_{XY} = - L_X\cdot\sin\alpha + L_Y\cdot\cos\alpha\\
 \displaystyle d_Z    =   L_Z - \frac{L_X\cdot{p_X} + L_Y\cdot{p_Y}}{\pt}\cdot\frac{p_Z}{\pt}.
\end{array}
\label{eq:ip}
\end{equation}

The momentum $\vec{p}$ is calculated as the vector sum of the \Pep\Pem~ pair momenta at the conversion vertex.

The \Pep\Pem~ tracks at the conversion vertex are produced with a small opening angle. As a kinematic constraint on their $\theta$ and $\phi$ angular directions, the momenta of the tracks are required to be parallel. This constraint is applied to select conversion tracks. Similarly, the conversion vertex is fitted using kinematic constraints. Three algorithms are used to reconstruct conversions: the ``tracker-only''~\cite{TRK-10-001, TRK-10-003}, the ``ECAL-seeded''~\cite{NOTE-2006-005}, and the ``Gaussian sum filter'' (GSF)~\cite{EGM-10-001}. The tracker-only algorithm selects track pairs from all reconstructed tracks, with the kinematic constraint, and fits the conversion vertex; the ECAL-seeded algorithm takes the energy deposits in the ECAL as seeds and then extrapolates back to the tracker to fit the conversion tracks and vertices; the GSF electron algorithm follows a procedure similar to the tracker-only algorithm but uses electrons identified by the GSF algorithm to find the track pairs. The conversions from these three algorithms are merged and the duplicates are removed. Among the reconstructed conversion tracks, the one with the largest number of hits is kept if it passes the necessary quality cut. To be selected, the tracks of two opposite-signed  conversion particles both need at least five valid hits; in addition, the conversion vertex requires a valid fit with $\chi^2$ probability $> 5\times 10^{-4}$.

\begin{figure}[h!]
  \begin{center}
    \includegraphics[width=0.64\textwidth]{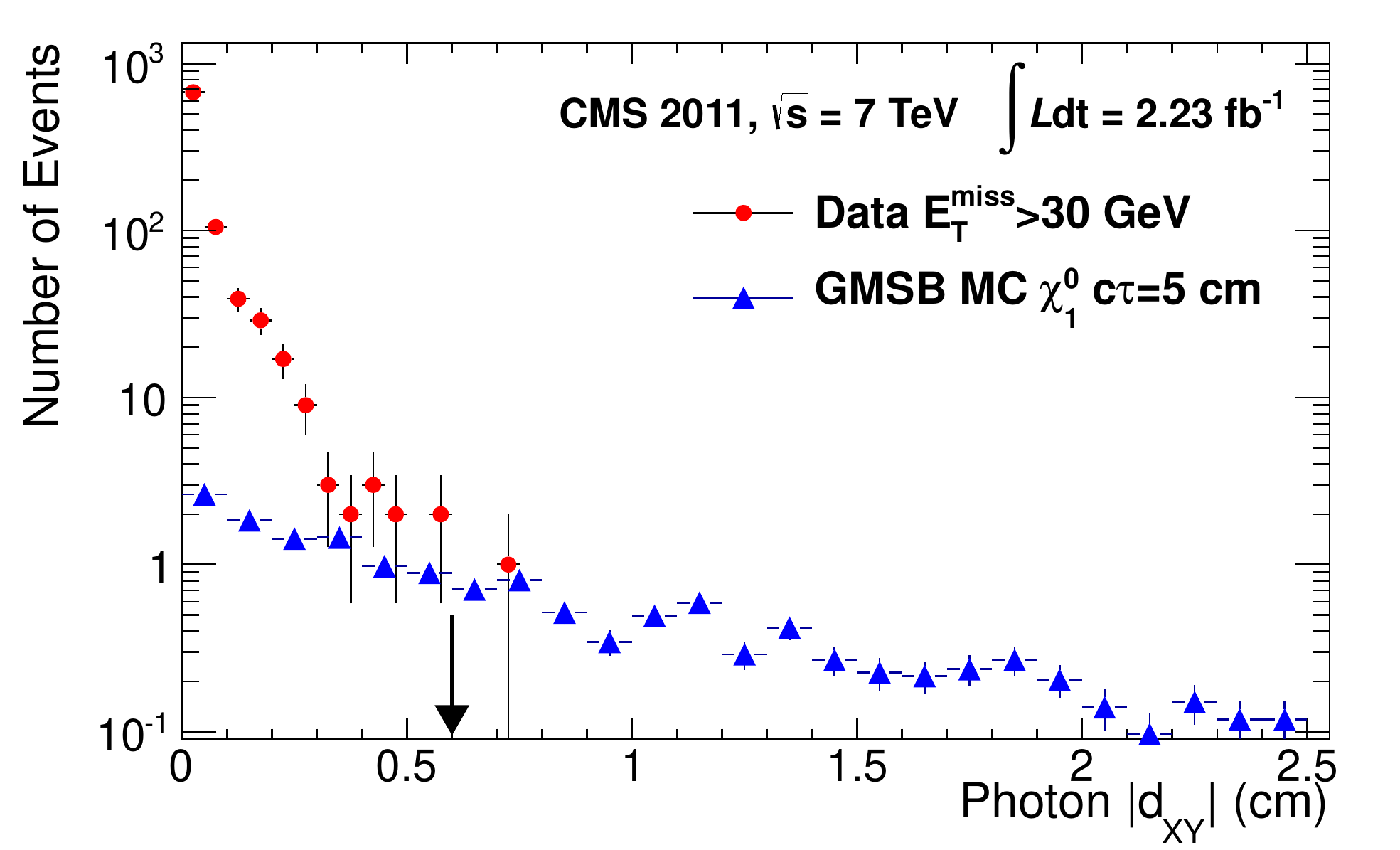}
    \caption{The $\abs{d_{XY}}$ distribution for data with $\MET > 30$\GeV compared with signal simulation for $c\tau$ = 5\unit{cm} normalized to the integrated luminosity of the data. The arrow indicates the $\abs{d_{XY}}$ requirement of the signal region.}
    \label{fig:ip_dxy_signal}
  \end{center}
\end{figure}

In high luminosity conditions, multiple collisions occur within a single bunch crossing, producing a number of primary vertices. The true primary vertex can have a large deviation from the interaction point in the longitudinal direction but a much smaller uncertainty in the transverse direction. To be robust against pile-up conditions, the transverse impact parameter ($d_{XY}$) with respect to the transverse position of the beam line is used in this analysis. To illustrate the $d_{XY}$ signature, the distribution of its absolute value in data for events with $\MET > 30 $\GeV is compared with a Monte Carlo sample with $\PSGczDo$ proper decay length $c\tau$ = 5\unit{cm}, as shown in Figure~\ref{fig:ip_dxy_signal}. The requirement that $\abs{d_{XY}}$ be larger than 0.6\unit{cm} then defines the signal region. This cut on $\abs{d_{XY}}$ is derived by optimizing the expected limits on the production cross section. This limit-based optimization is cross-checked against an optimization on the expected signal significance, and the results obtained are the same. In addition, we evaluate the search for several values of the $\abs{d_{XY}}$ cut, in the range from 0 to 1.0 cm, and establish that the results are stable for both expected and observed limits. As seen in Figure~\ref{fig:ip_dxy_signal}, one event passes all selection criteria. The Monte Carlo sample is normalized to the integrated luminosity, and on average 8.3 events from the SPS8 benchmark model are expected to pass all selections.

The effect of the event selection is illustrated in Table \ref{tab:cut_flow} for the $c\tau$ = 5\unit{cm} sample. Of the 45\,057 simulated events, 711 events survive the cuts, for an overall event selection efficiency of 1.58\%. Efficiencies for four neutralino lifetimes are given in Table \ref{tab:eff_lifetime}. On average, the signal efficiency is about 1\% to 2\% after the trigger requirements and selection criteria are applied. Beyond $c\tau$ = 25\unit{cm}, the signal efficiency decreases due to the limitations of the tracking algorithms for reconstructing highly displaced tracks.
\begin{center}
\begin{table}[h!]
	  \topcaption{Signal selection cut flow for c$\tau$ = 5\unit{cm}.}
	\begin{center}
	\begin{tabular}{ m{7cm} | c }
		\hline
		Selection & Events in Monte Carlo\\
		\hline
		\hline
		Total & 45057\\
		\hline
		Diphoton trigger & 39988\\
		\hline
		Leading (sub-leading) photon \par $\et>45$ (30)\GeV & 37398\\
		\hline
		Any ECAL barrel photon $\et>45$\GeV \par and photon identification & 27766\\
		\hline
		Jets $p_{\mathrm{T}1} > 80$\GeV and $p_{\mathrm{T}2} > 50$\GeV & 26229\\
		\hline
		Conversion selection & 1602\\
		\hline
		$\MET > 30$\GeV & 1542\\
		\hline
		$\abs{d_{XY}}> 0.6$\unit{cm} & 711\\
		\hline
	\end{tabular}
	\end{center}
	\label{tab:cut_flow}
\end{table}
\end{center}

\begin{center}
\begin{table}[h!]
	\topcaption{Event selection efficiency \vs $\PSGczDo$ proper decay length.}
	\begin{center}
	\begin{tabular}{ l || c  c  c  c }
		\hline
		$c\tau$ [cm] & 2 & 5 & 10 & 25\\ \hline
		Efficiency & $0.92\%$ & $1.58\%$ & $1.80\%$& $1.39\%$ \\ \hline
		Statistical errors & $0.05\%$ & $0.06\%$ & $0.06\%$& $0.06\%$ \\ \hline
	\end{tabular}
	\label{tab:eff_lifetime}
	\end{center}
\end{table}
\end{center}

\section{Background determination}
The background contributions to the sample of events with photon-plus-jets with missing transverse energy arise mainly from QCD events with real photons, and QCD multijet events, due to the effect of finite detector resolution on the measurement of \MET. In QCD events with real photons, there are typically one real photon and one jet that is misidentified as a photon in the final selection. The rate of QCD multijet events is sufficiently large to have a finite probability for jets to be misidentified as photons. There are additional sources of background with intrinsic missing transverse energy, such as the W+jets process, but these are heavily suppressed by the \et cut for the photon and make a negligible contribution to the final selection. The $\PSGczDo\rightarrow\gamma +\PXXSG$ decay has two signatures: \MET from the unseen $\PXXSG$ and a large impact parameter from the displaced photons. The strategy for determining the background is to use control samples that are kinematically similar to the candidate sample while having no real \MET. The data with $\MET < 20$\GeV are selected as such a control sample. The background is then estimated as the number of events with $\MET < 20$\GeV and satisfying all the other selection criteria.

\begin{figure}[h!]
  \begin{center}
    \includegraphics[width=0.64\textwidth]{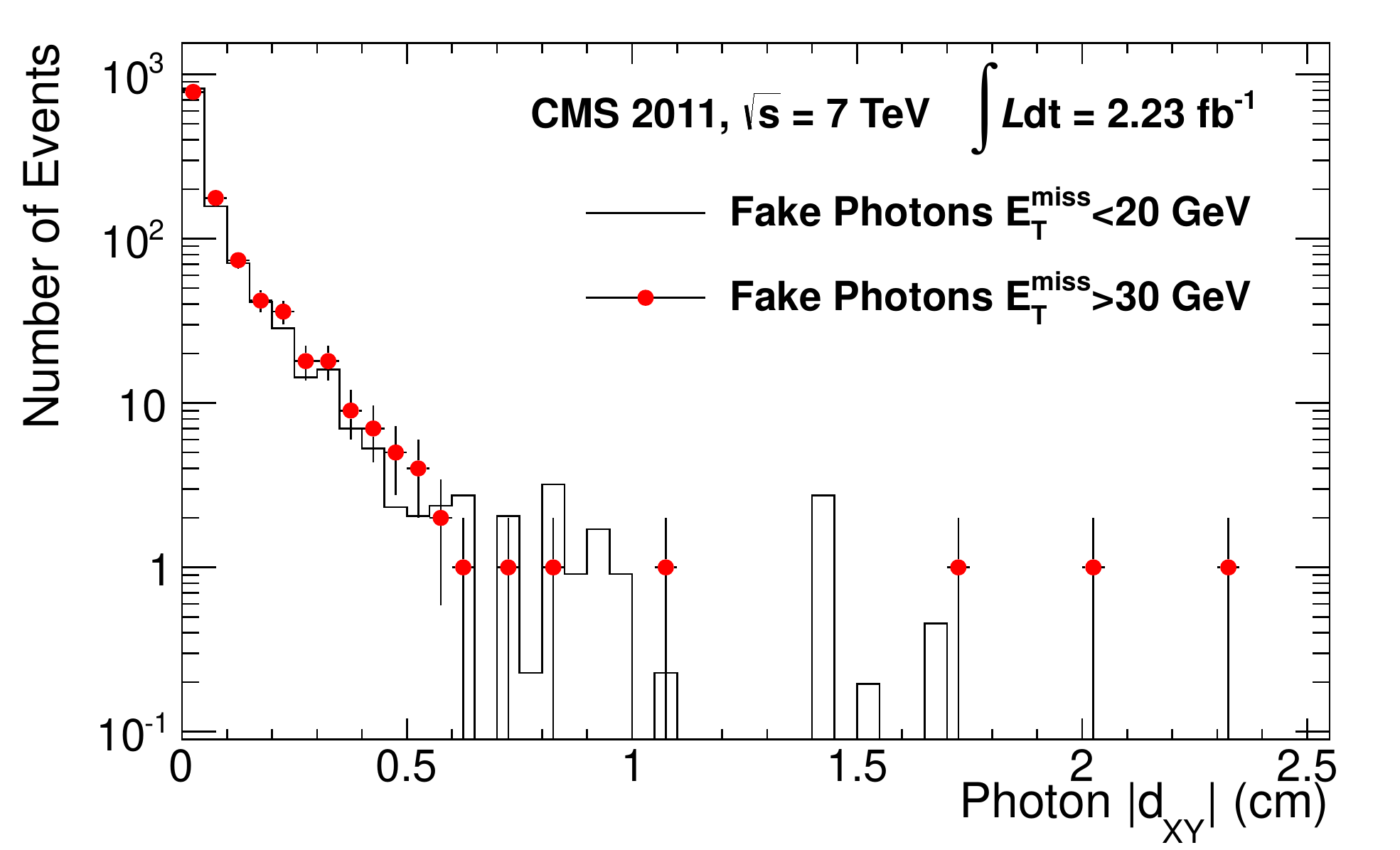}
    \includegraphics[width=0.64\textwidth]{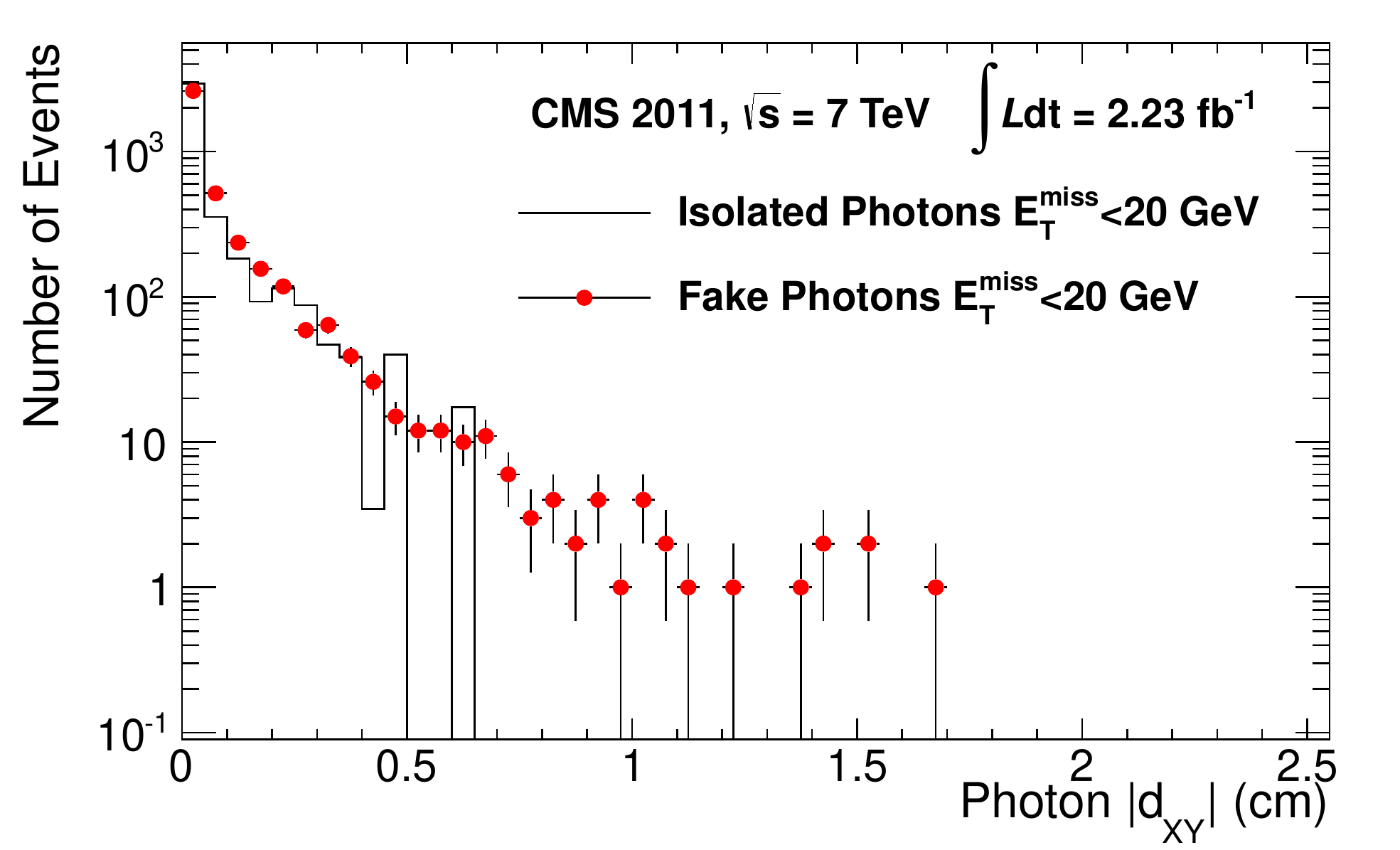}
	\caption{Top: Fake photon $\abs{d_{XY}}$ distributions in the background ($\MET < 20$\GeV) and signal ($\MET > 30$\GeV) regions. Bottom: Isolated and fake photon $\abs{d_{XY}}$ distributions in the background region ($\MET < 20$\GeV).}
    \label{fig:ip_dxy_real_fake}
  \end{center}
\end{figure}

To ensure that the $\MET < 20 $\GeV region of data gives a correct estimation of the background, the $\abs{d_{XY}}$ distributions of fake photons, defined previously, and isolated photons are compared. First, the $\abs{d_{XY}}$ distribution of fake photons for events with $\MET < 20$\GeV (background region) is normalized by the total number of conversions and compared with the $\abs{d_{XY}}$ distribution of fake photons in the region where $\MET > 30$\GeV (signal region), as shown in Figure~\ref{fig:ip_dxy_real_fake}(top). The $\abs{d_{XY}}$ distribution in the $\MET < 20$\GeV (background) region is reweighted by the conversion vertex $\chi^2$ probability of the $\MET < 20$\GeV (signal) region. This takes into account that the distributions of the $\chi^2$ probability are not the same in the background and signal region. Second, the $\abs{d_{XY}}$ distributions of isolated photons and fake photons in the low \MET region are compared by normalizing to the total number of conversions, also reweighted by the conversion vertex $\chi^2$ probability, as shown in Figure~\ref{fig:ip_dxy_real_fake}(bottom). The $\abs{d_{XY}}$ distributions of fake photons and isolated photons agree, and thus we conclude that the $\abs{d_{XY}}$ distribution for events with $\MET < 20$\GeV gives a good description of the background.  The residual differences in the fake and isolated photon distributions in the low \MET control sample are consistent with the expected enhancement of jets misreconstructed as photons in the fake photon sample. We therefore take the isolated photon sample in the low \MET control sample as the basis for the background estimate in the signal region, after we apply conversion vertex $\chi^2$ probability reweighting.

\begin{figure}[h!]
  \begin{center}
    \includegraphics[width=0.64\textwidth]{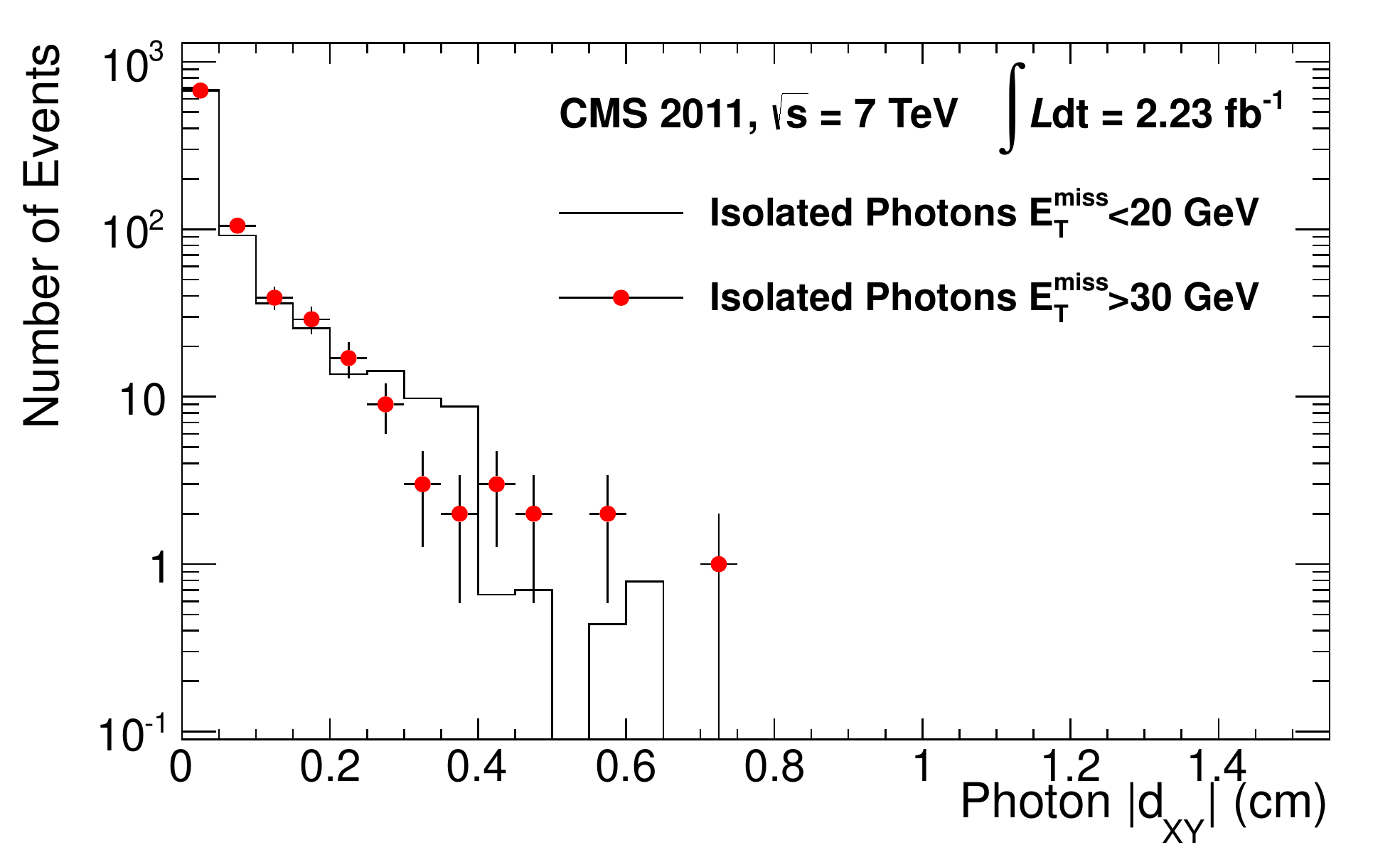}
    \caption{$\abs{d_{XY}}$ distributions for isolated photons: background region ($\MET < 20$\GeV) compared to the signal region ($\MET > 30$\GeV).}
    \label{fig:ip_dxy_linear}
  \end{center}
\end{figure}

The isolated photon $\abs{d_{XY}}$ distributions for the signal and background regions are shown in Figure~\ref{fig:ip_dxy_linear}. They are normalized to the same number of conversions and reweighted by the conversion vertex $\chi^2$ probability to predict the number of background events. A requirement that $\abs{d_{XY}} > 0.6$\unit{cm} is applied to the background region $\MET < 20$\GeV, which gives a total background of $0.78_{-0.48}^{+1.25}$ events. As a cross check, Monte Carlo background samples are fitted with a general functional form to extrapolate the expected tail of events for isolated photons beyond the $\abs{d_{XY}} > 0.6$\unit{cm} cut. Applying this general functional form to the background sample derived from data and integrating the expected rate above the $\abs{d_{XY}} > 0.6$\unit{cm} cut gives a background estimate consistent with the number quoted above.

\section{Systematic uncertainties}
Table ~\ref{tab:sus-uncertainty} summarizes the systematic uncertainties affecting this analysis. To determine the uncertainty in the conversion reconstruction efficiency, we compared the predicted and observed numbers of conversions for $Z\rightarrow\mu\mu\gamma$ events. For photons with $\et>20$\GeV, a scale factor of $0.87\pm0.08$ is derived. High \et photons have a reconstruction efficiency of $(7.0\pm0.5)\%$ in the Drell--Yan to dimuon Monte Carlo samples, and the corresponding high \et photon efficiency in the GMSB Monte Carlo samples is $(6.0\pm0.7)\%$, yielding a $\cPZ\rightarrow\Pgm\Pgm\Pgg$--GMSB scale factor of $1.16\pm0.08$. Considering the correlations between the conversion reconstruction efficiency and the high-\et dependence, the uncertainty of conversion reconstruction is determined to be 21\% for the signal region.

The $d_{XY}$ resolution is studied by comparing the $d_{XY}$ resolution for converted photons from $\cPZ\rightarrow\Pgm\Pgm\Pgg$ events between Monte Carlo and data. The data and Monte Carlo samples are in agreement and a $d_{XY}$ resolution of 0.06\unit{cm} extracted from data gives less than $0.5\%$ uncertainty for the signal selection.  The resolution uncertainty includes tails in the fitted resolution function.

Other sources of uncertainty include the integrated luminosity (2.2\%), the jet \pt/\MET selection requirement, and photon identification. The jet \pt/\MET resolution contributes less than $0.5\%$ to the total uncertainty~\cite{JME-10-003, JME-10-009}. The photon identification efficiency is studied in \cite{SUS-10-002} and includes the uncertainties from pile-up (2.5\%), the photon data/MC scale measured in a $\cPZ\to \Pe\Pe$ data sample (2.6\%), and the photon--electron identification difference as studied in MC samples (0.5\%). Other sources of uncertainty give negligible contributions to the systematic uncertainty. The systematic error from the $d_{XY}$ cut is taken into account when evaluating the limits on the contribution from the photon $d_{XY}$ resolution. The total systematic uncertainty is determined to be 25\% by considering all the correlations among pile-up, photon identification data/MC scale, and conversion reconstruction efficiency. 

\begin{center}
\begin{table}[h!]
	\topcaption{Summary of systematic uncertainties.}
	\begin{center}
	\begin{tabular}{ l || c }
		\hline
		Systematics & Uncertainty ($\%$)\\
		\hline
		Integrated luminosity & 2.2\\
		Jet \pt/\MET energy scale & $<0.5$ \\
		Pile-up & 2.5\\
		Photon identification data/MC scale & 2.6\\
		Photon--electron difference & 0.5\\
		Conversion reconstruction efficiency & 21\\
		Photon $d_{XY}$ resolution & $<0.5$\\
		Total & 25 \\
		\hline
	\end{tabular}
	\end{center}
	\label{tab:sus-uncertainty}
\end{table}
\end{center}

\section{Results and interpretation}
\begin{figure}[h!]
  \begin{center}
    \includegraphics[width=0.64\textwidth]{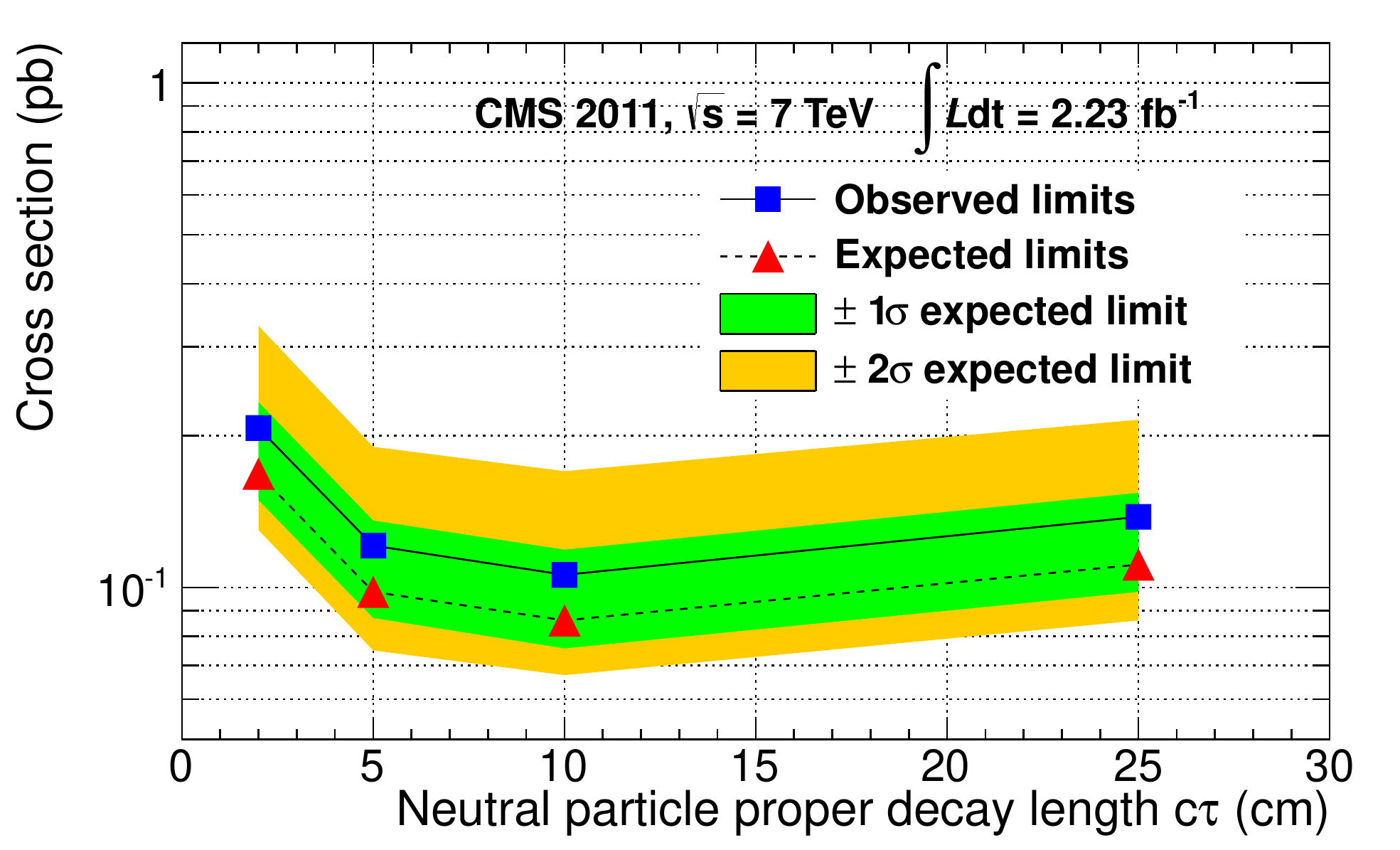}
    \caption{$95\%$ confidence level upper limits on the pair production cross section for neutral particles, each of which decays into a photon and invisible particles, as a function of the neutral particle proper decay length. The observed values as a function of the neutral particle proper decay length are shown by the solid line. The dashed line indicates the expected median of results for the background-only hypothesis, while the green (dark) and yellow (light) bands indicate the ranges that are expected to contain $68\%$ and $95\%$ of all observed excursions from the median, respectively.}
    \label{fig:upper_limit}
  \end{center}
\end{figure}

One event with $\abs{d_{XY}} = 0.74$\unit{cm} and \MET = 44.9\GeV satisfying all the other selection criteria is observed. The estimated background is $0.78_{-0.48}^{+1.25}$ events. We determine the upper limits for the cross section for pair production of neutral particles, each of which decays into one photon and invisible particles. A CLs limit setting method~\cite{Read:2002hq, Moneta:2010pm} is employed using log-normal uncertainties for the total background rate to incorporate the uncertainties in the total background rate, integrated luminosity, and total acceptance times efficiency. The observed $95\%$ confidence level limits vary between 0.11 and 0.21\unit{pb}, depending on the neutral particle proper decay length (Table \ref{tab:upper_limit} and Figure~\ref{fig:upper_limit}).

\begin{table}[h!]
\begin{center}
	\topcaption{$95\%$ confidence level (CL) upper limits on the cross section for pair production of neutral particles, each of which decays into a photon and invisible particles, as a function of the neutral particle proper decay length.}
	\begin{center}
	\begin{tabular}{ c || c  c  c  c }
		\hline
		$c\tau$ (cm) & 2 & 5 & 10 & 25\\ \hline
		$\sigma$ (pb) $95\%$ CL & 0.21 & 0.12 & 0.11 & 0.14 \\ \hline 
	\end{tabular}
	\end{center}
	\label{tab:upper_limit}
\end{center}
\end{table}

\section{Summary}
We have introduced a novel method using the photon conversion impact parameter to search for new physics involving long-lived particles decaying into photons. The high resolution of the impact parameter reconstruction enabled our analysis to probe for new physics to much lower values of \MET compared to other LHC searches~\cite{SUS-10-002}. The search was performed using the final state of photons, jets, and missing transverse energy. Cross-section limits on pair production for such particles, each of which decays into a photon and invisible particles, were set as a function of the long-lived particle's lifetime. These upper limits are applicable to a general class of new physics processes with diphotons in the final state.

\section*{Acknowledgments}
We congratulate our colleagues in the CERN accelerator departments for the excellent performance of the LHC machine. We thank the technical and administrative staff at CERN and other CMS institutes, and acknowledge support from: FMSR (Austria); FNRS and FWO (Belgium); CNPq, CAPES, FAPERJ, and FAPESP (Brazil); MES (Bulgaria); CERN; CAS, MoST, and NSFC (China); COLCIENCIAS (Colombia); MSES (Croatia); RPF (Cyprus); MoER, SF0690030s09 and ERDF (Estonia); Academy of Finland, MEC, and HIP (Finland); CEA and CNRS/IN2P3 (France); BMBF, DFG, and HGF (Germany); GSRT (Greece); OTKA and NKTH (Hungary); DAE and DST (India); IPM (Iran); SFI (Ireland); INFN (Italy); NRF and WCU (Korea); LAS (Lithuania); CINVESTAV, CONACYT, SEP, and UASLP-FAI (Mexico); MSI (New Zealand); PAEC (Pakistan); MSHE and NSC (Poland); FCT (Portugal); JINR (Armenia, Belarus, Georgia, Ukraine, Uzbekistan); MON, RosAtom, RAS and RFBR (Russia); MSTD (Serbia); MICINN and CPAN (Spain); Swiss Funding Agencies (Switzerland); NSC (Taipei); TUBITAK and TAEK (Turkey); STFC (United Kingdom); DOE and NSF (USA).
\bibliography{auto_generated}   

\cleardoublepage \appendix\section{The CMS Collaboration \label{app:collab}}\begin{sloppypar}\hyphenpenalty=5000\widowpenalty=500\clubpenalty=5000\textbf{Yerevan Physics Institute,  Yerevan,  Armenia}\\*[0pt]
S.~Chatrchyan, V.~Khachatryan, A.M.~Sirunyan, A.~Tumasyan
\vskip\cmsinstskip
\textbf{Institut f\"{u}r Hochenergiephysik der OeAW,  Wien,  Austria}\\*[0pt]
W.~Adam, E.~Aguilo, T.~Bergauer, M.~Dragicevic, J.~Er\"{o}, C.~Fabjan\cmsAuthorMark{1}, M.~Friedl, R.~Fr\"{u}hwirth\cmsAuthorMark{1}, V.M.~Ghete, J.~Hammer, N.~H\"{o}rmann, J.~Hrubec, M.~Jeitler\cmsAuthorMark{1}, W.~Kiesenhofer, V.~Kn\"{u}nz, M.~Krammer\cmsAuthorMark{1}, I.~Kr\"{a}tschmer, D.~Liko, I.~Mikulec, M.~Pernicka$^{\textrm{\dag}}$, B.~Rahbaran, C.~Rohringer, H.~Rohringer, R.~Sch\"{o}fbeck, J.~Strauss, A.~Taurok, W.~Waltenberger, G.~Walzel, E.~Widl, C.-E.~Wulz\cmsAuthorMark{1}
% ...
\vskip\cmsinstskip
\dag:~Deceased\\
1:~~Also at Vienna University of Technology, Vienna, Austria\\
% ...

\end{sloppypar}
\end{document}